# On the connection between weak measurement in quantum physics and analytic phase-retrieval in classical wave optics

**Nobuharu Nakajima**

Formerly at the Department of Engineering, Shizuoka University, 3-5-1 Johoku, Chuou-ku, Hamamatsu 432-8561, Japan

e-mail: nakajima.nobuharu@shizuoka.ac.jp

Abstract

The physical interpretation of weak measurements has been the subject of much debate. It is known that anomalous phenomena and results that appear in weak measurements are essentially related to the phase of the quantum system being measured. Consideration of the phase is important to clarify its physical interpretation. In classical wave optics, there has long been studies on methods of measuring or retrieving the phase of a wave function. We here present that one of those methods, the analytic phase retrieval based on the properties of entire functions, has a close connection with weak measurements in quantum physics. We explain such a connection for two emblematic optical weak-measurements that have the same mathematical formalism as quantum systems: one is a system for weak measurements of polarized light displacement in a birefringent crystal, and the other is a system for the direct measurement of a wave function by weakly coupling it to a pointer. In those two systems, we show that the pre- and post-selection of polarized light provides a filtering effect similar to that utilized in the analytic phase retrieval.

# I.    INTRODUCTION

In 1988, Aharonov, Albert, and Vaidman [1] derived that the usual measuring procedure for preselected and postselected ensembles of quantum systems produces unusual results under some natural conditions of weakness of the measurement, the so-called weak measurement. In such a measurement system, they defined a new concept: a weak value of a quantum variable. The weak value has the property that it has a large value under certain measurement conditions, thereby producing a very large signal amplification. Furthermore, unlike the expectation values of ordinary quantum variables, weak values are complex numbers in general. There properties have led to various developments concerning weak values [2-5]. Especially in the 2000s, experiments using polarization of light [6-11], which has the same mathematical form as quantum systems with two states, have shown some applications of weak values and have attracted widespread attentions [12].

The anomalous behavior of the measurements with weak values is due to interference effects caused by the phase of a quantum state changed by a pre- and post-selected system [1]. In addition, it has been shown that the abstract theoretical quantity such as a wave function in quantum physics can also be expressed in terms of complex weak values [11], and measuring both the real and imaginary parts of a complex weak value is equivalent to retrieving the phase of the quantum state. Therefore, in weak measurements, the phase of the quantum state being measured is the most important factor [13].

The problem of determining the quantum state, i.e., the phase retrieval problem in quantum physics, has long been known as the Pauri problem [14-18], which began with Pauri's question of whether knowledge of its probability distributions of position and momentum would be sufficient to determine the wave function of a particle. The problem of phase retrieval from the absolute square of a wave function is also important in classical physics. The study of such phase retrieval has long been conducted in the field of structure determination of objects using light, x-ray, and electron microscopy [19].

In the field of classical wave optics, the standard method of phase recover is interferometry or holography: that is, the modulus and phase of an unknown wave function are recovered from interference fringes obtained by coherently adding a known reference wave function to the unknown wave one. There have also been studies of another approach to the phase retrieval method, in which the phase of the unknown wave function is retrieved from its some intensity distributions without such a coherent reference wave. The study of this type of approach is useful in measurement systems where it is inconvenient to generate the reference wave, for example, when using x-rays, electron waves, or atomic waves. Various methods have been proposed for this type of phase-retrieval approach, including iterative methods based on computer algorithms [20-22], analytic methods based on the properties of entire functions [23-26], and methods using solutions of the transport-of-intensity equation for waves [27-29]. Some of these phase-retrieval methods in classical wave optics have been shown to be applicable to the reconstruction of wave functions in quantum systems (see [30] and references therein).

We here present that weak measurements are closely related to the analytic phase retrieval based on the properties of entire functions [23-26]. In particular, we focus on two emblematic cases of weak measurements: the large signal amplification by a weak value in the system of spin-1/2 particles [1] and the weak measurement of the real and imaginary parts of a wave function [11]. In the analytic phase retrieval, the phase of a one-dimensional wave function is determined [23] by the positions of zeros in the complex plane of the Fourier transform function of the wave function, and its Fourier phase can be retrieved analytically from two Fourier intensities of the exponential filtered and unfiltered wave functions [25, 26]. Thus, we first show that the anomalous displacement of probe wave packets in measurements of the spin Hall effect of polarized light [10], mathematically analogous to measurement systems using spin-1/2 particles [1], can be easily explained by the effect of a zero in the complex plane obtained by expanding the momentum space. Next, we reveal that the weak measurement of a wave function [11] is essentially based on the use of an amplitude or phase modulation filter brought about by pre- and post-selections of polarized light, and that it is a filtering-based method similar to the analytic phase retrieval method with an exponential filter [25, 26].

This paper is organized as follows. In Sec. II, we review the analytic phase-retrieval method based on the properties of entire functions. In Sec. III, we show that the anomalous displacement of probe beams in the measurement system of polarized light with a birefringent crystal can be easily interpreted as an effect on the phase due to a zero in the complex plane. In Sec. IV, we present that the weak measurement of a wave function in the two-state system with polarized light belongs to the filtering-based reconstruction method similar to the analytic phase retrieval method. Concluding remarks are given in Section V.

## II. Phase retrieval based on the properties of entire functions

In this section, we review the analytic phase-retrieval method for reconstructing wave functions by use of the mathematical properties of entire functions of exponential type. We first consider a Fourier-transforming system with some optical components such as convex lens as shown in Fig. 1. For simplicity, a one-dimensional object of complex-amplitude transmittance $f(x)$ [with a finite extent of interval $(a, b)$] is assumed here, where the object plane is defined as the plane immediately behind the object perpendicular to the optical axis. When the object is illuminated by a coherent monochromatic plane wave of unit amplitude, the complex amplitude function $F(p)$ in the Fourier plane is given by

$$F(p) = \int_a^b f(x) \exp(-ixp)\, dx, \tag{1}$$

where multiplicative constants associated with the diffraction integrals are ignored since these are not essential to discussions in this paper.

By changing the real variable $p$ to the complex one with $z = p + iq$ in Eq. (1), the function $F(p)$ on the real axis $p$ is extended to the function in the complex plane. Then the function $F(z)$ becomes an entire function from a theorem formulated originally by Paley and Wiener [31]. It is well known that an entire function of finite order such as $F(z)$ may be described everywhere (to within a linear phase factor) by its zeros with the expression being known as a Hadamard product,

$$F(z) = z^d \exp[i(a+b)z] B \prod_{j=-\infty}^{\infty} \left(1 - \frac{z}{z_j}\right), \tag{2}$$

where $B$ is a scaling constant, $d$ is the order of a zero at the origin, and $z_j$ is the vector notation of the $j$-th zero of $|F(z)|$ in the complex plane. If the location of zeros in the complex plane could be determined from the modulus $|F(p)|$, the complex amplitude function $F(p)$ would be obtained (i.e., the phase of $F(p)$ could be retrieved). However, it is impossible [23] to determine the location of zeros in the complex plane from a single observable intensity distribution $|F(p)|^2$.

As an alternative to locating zeros, a linear method using an exponential filter has been proposed to retrieve the phase of the complex amplitude function from two intensity distributions in the Fourier plane [25]. One is a Fourier-intensity distribution of $|F(p)|^2$, and the other is a Fourier-intensity distribution $|\tilde{F}(p)|^2$ of the object function modulated by an exponential function $\exp[-c(x-s)]$, where $c$ is a known constant and $s$ is a known constant smaller than $a$ in Eq. (1). Then the function $\tilde{F}(p)$ is given by

$$\tilde{F}(p) = \int_a^b f(x) \exp[-c(x-s)]\exp(-ipx)\, dx$$

$$= \exp(cs) \int_a^b f(x) \exp[-ix(p-ic)]dx$$

$$= \exp(cs) F(p-ic). \tag{3}$$

Since the amplitude transmittance of the exponential filter over the range of that object function is less than unity, this filter can be easily realized as an attenuation filter.

For the derivation of the phase retrieval equation, we first write the function $F(p)$ as

$$F(p) = M(p)\exp[i\phi(p)], \tag{4}$$

where $M(p)$ and $\phi(p)$ denote the Fourier modulus $|F(p)|$ and the phase of $F(p)$, respectively. Expanding the real variable $p$ into the complex one, $p - ic$, in Eq. (4), the complex function along the line for $q = -c$ in the complex plane is given by

$$F(p-ic) = M(p-ic)\exp[i\text{Re}\phi(p-ic) - \text{Im}\phi(p-ic)], \tag{5}$$

where Re and Im denote the real and imaginary parts of the phase function $\phi(p-ic)$ extended in the complex plane. Substituting Eq. (5) into Eq. (3), we can obtain the relation between observable moduli $|F(p)|$ and $|\tilde{F}(p)|$,

$$\frac{|\tilde{F}(p)|}{|M(p-ic)|} = \exp[cs - \text{Im } \phi(p-ic)], \tag{6}$$

where

$$|M(p-ic)| = \sqrt{F(p-ic)F^*(p+ic)}, \tag{7}$$

in which the asterisk denotes a complex conjugate. The left-hand term of Eq. (6) can be calculated from the observable intensity distributions because $|\tilde{F}(p)|$ is the square root of the Fourier intensity of the filtered object in Eq. (3) and because $|M(p-ic)|$ is the modulus of the Fourier transform of the product of the inverse Fourier transform of $M(p)$ and the known exponential function $\exp(-cx)$ [25].

Equation (6) can be solved numerically using the Fourier series expansion representation of the phase $\phi(p)$ [25], but an analytic solution can be derived [32] as shown in Appendix A:

$$\phi(p) = \Im\left\{\frac{\Im^{-1}[D(p',c)]}{-i\sinh(cx)}\right\} \quad (x \neq 0), \tag{8}$$

where $D(p',c) = \ln[|\tilde{F}(p')|/|M(p'-ic)|] - cs$, and $\Im\{\cdots\}$ and $\Im^{-1}[\cdots]$ denote the Fourier and the inverse Fourier transforms, respectively. As can be seen from the right-hand side of Eq. (6), the value of the inverse Fourier transform of $D(p,c)$ at the origin of $x = 0$ is concerned with the inclination of the linear phase factor of $\phi(p)$. The effect of such a linear phase in the Fourier plane corresponds to a shift of the reconstructed object function. Since the exact position of an object function is not so important for many applications, the influence of can be ignored in practice. Therefore, the term related to a linear phase is omitted here for simplicity, but could be incorporated into Eq. (8) as described in Appendix A. Consequently, the object wave function $f(x)$ can be reconstructed by taking the inverse Fourier transform of the complex amplitude function $F(p)$ consisting of the measured modulus $M(p)$ and the retrieved phase $\phi(p)$.

The present phase-retrieval method is applicable for all kinds of functions except Hermitian object functions [i.e., $f(x) = f^*(-x)$]. To reconstruct such an Hermitian object, the logarithmic Hilbert transform has to be used in the phase retrieval together with the present method [26]. In addition, we have presented an extension of the present one-dimensional method to two-dimensional phase retrieval, which has been demonstrated by optical experiments using films with exponential amplitude transmittance [33].

As shown in Eqs. (5) and (6), the phase of $M(p-ic)$ and the real part $\text{Re}\phi(p-ic)$ have no effect on the intensity measurements in the Fourier plane. In the image plane of Fig. 1, however, they lead to a transverse displacement of the object function $f(x)$ modulated by the exponential filter. This is closely related to the large displacement in weak measurements.

This connection is discussed in Sec. III. Furthermore, in Sec. IV, we discuss the relationship of this phase retrieval method to the measurement method of the transverse wave function of light utilizing the concept of weak measurements.

## III. Interpretation of anomalous probe displacement in weak measurements by properties of entire functions

In this section, we consider the symbolic phenomenon of probe displacement anomalies in weak measurements, which is originally addressed in a two-state system of spin-1/2 particles by Aharonov *et al.* [1]. This phenomenon can be alternatively realized by the spin Hall effect of polarized light [10], because such systems have the same mathematical form regardless of quantum or classical theory. Thus, we discuss the relation between the anomalous probe displacement and the properties of entire functions in the complex plane by use of the optical system with a birefringent crystal in Fig. 2.

### A. Anomalous probe displacement in the usual formulation

In Fig. 2, the initial state $|\psi_i\rangle$ of a probe beam after passing through the polarizer $P_1$ is set to $|\psi_i\rangle = |\psi_0\rangle|S\rangle$, where $|\psi_0\rangle$ represents the transverse beam profile, and $|S\rangle$ is a pre-selected polarization state of $|S\rangle = (|H\rangle + |V\rangle)/\sqrt{2}$, in which $|H\rangle$ and $|V\rangle$ denote the horizontal and vertical polarization states, respectively. The wave function of the initial state in the transverse position $x$ is given by

$$\langle x|\psi_i\rangle = \langle x|\psi_0\rangle|S\rangle = \psi_0(x)|S\rangle. \tag{9}$$

The phase distribution of the probe beam $\psi_0(x)$ is assumed here to be constant, and thus the probe beam is centered at the origin in the transverse momentum space $p$, conjugate to the position $x$. We use here an interaction Hamiltonian $\hat{H}_I = g\hat{A}\otimes\hat{p}$, where $g$ is a coupling constant, $\hat{A}$ is an observable, and $\hat{p}$ is the transverse momentum operator.

When one of the Pauli matrices $\hat{\sigma}_z = |H\rangle\langle H| - |V\rangle\langle V|$ is used as the observable $\hat{A}$, the output state of the probe beam after passing through the birefringent crystal at an interaction time period $\tau$ can be written as

$$|\psi_1\rangle = \exp(-i\epsilon\hat{\sigma}_z p)|\psi_i\rangle$$
$$= [\cos(\epsilon p)|S\rangle - i\sin(\epsilon p)|D\rangle]|\psi_0\rangle, \qquad (10)$$

where $|D\rangle$ denotes the polarization state of $|D\rangle = (|H\rangle - |V\rangle)/\sqrt{2}$, orthogonal to $|S\rangle$, and $\epsilon = g\tau$ is the displacement of each polarization component, of which the absolute value is sufficiently smaller than the extent of probe beam $\psi_0(x)$ (weak measurement)

Then the post-selected polarization state $|D_\theta\rangle$ of the polarizer P$_2$ in Fig. 2 is defined by

$$|D_\theta\rangle = \sin\theta|S\rangle + \cos\theta|D\rangle, \qquad (11)$$

where the absolute value of $\theta$ is generally less than about $\pi/9$ [10, 11], and thus $|D_\theta\rangle$ is a polarization state close to the state $|D\rangle$ [e.g., $\sin(\pi/10) = 0.30902$, $\cos(\pi/10) = 0.95106$]. The final state of light passing through the polarizer P$_2$ just before entering a charge coupled device (CCD) detector is given by

$$|\psi_f\rangle = \langle D_\theta|\psi_1\rangle = [\cos(\epsilon p)\sin\theta - i\sin(\epsilon p)\cos\theta]|\psi_0\rangle$$
$$= A_\theta(p)\exp[i\Phi_\theta(p)]|\psi_0\rangle, \qquad (12)$$

where

$$A_\theta(p) = |\cos(\epsilon p)\cos\theta|\sqrt{\tan^2\theta + \tan^2(\epsilon p)}, \qquad (13)$$

and

$$\Phi_\theta(p) = -\tan^{-1}\left[\frac{\tan(\epsilon p)}{\tan\theta}\right]. \qquad (14)$$

Under the weak measurement condition that $|\epsilon p| \ll 1$, Eq. (12) can be approximated as

$$|\psi_f\rangle \cong |\sin\theta| \exp\left[-i\frac{\epsilon p}{\tan\theta}\right]|\psi_0\rangle. \tag{15}$$

Thus, this final state of light in the CCD detector, expressed in the $x$ representation, is given by

$$\langle x|\psi_f\rangle \cong |\sin\theta|\,\psi_0\left(x - \frac{\epsilon}{\tan\theta}\right). \tag{16}$$

In this weak measurement, the weak value $\langle \hat{A} \rangle_w$ can be obtained from Eqs. (10) and (11) as

$$\langle \hat{A} \rangle_w = \frac{\langle D_\theta|\hat{A}|S\rangle}{\langle D_\theta|S\rangle} = \frac{1}{\tan\theta}. \tag{17}$$

When $|\epsilon \Delta p| \ll |\theta| \ll 1$ (where $\Delta p$ is the momentum variance of the probe) [13], the large displacement of the probe beam in the transverse position space can be obtained, and then Eq. (16) can be further approximated by

$$\langle x|\psi_0\rangle \cong |\theta|\,\psi_0\left(x - \frac{\epsilon}{\theta}\right). \tag{18}$$

This phenomenon is caused by an abrupt gradient of the phase around $p = 0$ in Eq. (15). Thus, in the next subsection, it is shown that this abrupt phase change can be simply explained by zeros of entire functions in the complex plane.

**B. Explanation of anomalous displacements using zeros of entire functions**

When the magnitude of the angle of the post-selected state $|D_\theta\rangle$ in Eq. (11) is sufficiently small (i.e., $|\theta| \ll 1$), Eq. (11) can be approximately rewritten as

$$|D_\theta\rangle \cong \theta|S\rangle + |D\rangle$$

$$= \frac{1}{\sqrt{2}}[(1+\theta)|H\rangle - (1-\theta)|V\rangle]$$

$$\cong \frac{1}{\sqrt{2}}\left[e^{\theta}|H\rangle - e^{-\theta}|V\rangle\right]. \tag{19}$$

On the other hand, the output state of the probe beam from the birefringent crystal $|\psi_1\rangle$, expressed in terms of $|H\rangle$ and $|V\rangle$, is given by

$$|\psi_1\rangle = \frac{1}{\sqrt{2}}\left[e^{-i\epsilon p}|H\rangle + e^{i\epsilon p}|V\rangle\right]|\psi_0\rangle. \tag{20}$$

The final state of light passing through the polarizer $P_2$ in Eq. (12) can be rewritten from Eqs. (19) and (20) as

$$|\psi_f\rangle = \langle D_\theta|\psi_1\rangle$$

$$\cong \frac{1}{2}\left[e^{-i(\epsilon p + i\theta)} - e^{i(\epsilon p + i\theta)}\right]|\psi_0\rangle$$

$$= -i\sin(\epsilon p + i\theta)|\psi_0\rangle. \tag{21}$$

This equation indicates that the function $A_\theta(p)\exp[i\Phi_\theta(p)]$ in Eq. (12) can be approximated to the function extended in the complex plane from the function $-i\sin(\epsilon p)$ of a final state $\langle D|\psi_1\rangle$ obtained from Eq. (10) by using the polarization filter with $|D\rangle$ state, which is orthogonal to the initial polarization state $|S\rangle$. In other words, the relation between the two final states of light in post-selected polarization states $|D\rangle$ and $|D_\theta\rangle$ is equivalent to the relation between $F(p)$ and $F(p - ic)$ in Eqs. (4) and (5), provided that $|\theta| \ll 1$.

Under the condition ($|\epsilon \Delta p| \ll |\theta| \ll 1$) for the anomalous probe displacement, the phase of the function $-i\sin(\epsilon p \pm i|\theta|)$ is determined solely by the effect of a zero at the origin of the function $-i\sin(\epsilon p)$ (i.e., a zero of the modulus $|\sin(\epsilon p)|$). The zero at the origin of $-i\sin(\epsilon p)$ is shift by $\mp i|\theta|/\epsilon$ in the complex plane, as viewed from the function $-i\sin(\epsilon p \pm i|\theta|)$. Therefore, using Eq. (2), the effect of the zero at $z_0 = \mp i|\theta|/\epsilon$ along the real axis $z = p$ in the complex plane can be obtained as

$$\left(1 - \frac{z}{z_0}\right) = \left(1 - \frac{\epsilon p}{\mp i|\theta|}\right)$$

$$\cong \exp\left(\mp i \frac{\epsilon p}{|\theta|}\right). \tag{22}$$

When $|\theta| \ll 1$, Eq. (22) has a linear phase with a very sharp slope around $p = 0$, the same as in Eq. (15). Consequently, it can be seen that the large displacement of the probe beam in the transverse position space in Eq. (18) can be explained as the effect of the zero near the origin in the complex plane.

Recently, in weak measurements of the spin Hall shift of a polarized Gaussian beam due to partial reflection at a dielectric interface, it has been shown [34] that the zero of the spatial response function of the system shifts in the complex plane by a post-selected polarization angle that is approximately orthogonal to the pre-selected state. This shift of the zero in the complex plane can also be attributed to the filtering effect of the analytic phase retrieval described here.

## IV. Relationship between the weak measurement and the analytic phase retrieval in wave function measurements

As a typical application of the concept of weak measurements, the direct measurement of a transverse wave function of a photonic ensemble has been presented by use of a scanning system with a silver of waveplate through the wave field [11]. In this section, we discuss the similarities and differences between this direct measurement method and the analytical phase-retrieval method in Sec. II. That is, we show that the former directly measures the wave function utilizing the same filtering effects as the latter, while the latter analytically reconstructs the overall wave function without using weak measurement approximations.

## A. Direct measurement of the wave function using polarization

Following Lundeen *et al.* [11], we review the experiment for the direct measurement of a wave function as shown in Fig. 3. A one-dimensional object of complex-amplitude transmittance $\psi(x) = \langle x|\psi\rangle = |\psi(x)|\exp[i\varphi(x)]$ is illuminated by a stream of coherent photons with vertical polarization $|V\rangle$. The full initial state of light in the plane immediately behind the object is given by

$$|\psi_i\rangle = |\psi\rangle|V\rangle = \int dx\, \psi(x)|x\rangle|V\rangle. \tag{23}$$

After the light passes through the object, the polarization of the small portion of light around a position $x$ is rotated $\theta$ by a small rectangular silver of a half-wave plate ($\lambda/2$ silver). We assume here that the width of the silver plate is sufficiently smaller than the variation of the object function $\psi(x)$. Then the full state is therefore represented by

$$|\psi_i'\rangle = \exp(-i\,\theta \hat{A}\otimes\hat{\sigma}_y)|\psi_i\rangle = \int dx\, \psi(x)\exp(-i\,\theta\hat{A}\otimes\hat{\sigma}_y)|x\rangle|V\rangle, \tag{24}$$

where the observable $\hat{A}$ is the position operator $\hat{\pi}_x = |x_s\rangle\langle x_s|$ for the silver plate, of which the width was omitted, and $\hat{\sigma}_y$ is another one of the Pauli matrices, $\hat{\sigma}_y = -i(|H\rangle\langle V| - |V\rangle\langle H|)$, denoting the rotation of the linear polarization state of incident photons.

Since $|\theta|$ is small (about 20° or less) for weak measurements, we use the first order expansion of $\exp(-i\,\theta\hat{A}\otimes\hat{\sigma}_y)$ in Eq. (24). Then the field state $|\psi_p\rangle$ at a transverse momentum $p$ in the Fourier-transform plane of Fig. 3 is given by

$$|\psi_p\rangle = \langle p|\psi_i'\rangle \cong \langle p|\psi\rangle|V\rangle - i\,\langle p|\hat{A}|\psi\rangle\theta\hat{\sigma}_y|V\rangle$$

$$= \langle p|\psi\rangle\Big[|V\rangle - \langle\hat{A}\rangle_w \theta|H\rangle\Big], \tag{25}$$

where $\langle\hat{A}\rangle_w$ is the corresponding weak value, expressed as

$$\langle\hat{A}\rangle_w = \frac{\langle p|\hat{A}|\psi\rangle}{\langle p|\psi\rangle} = \frac{\langle p|x_s\rangle\langle x_s|\psi\rangle}{\langle p|\psi\rangle}$$

$$= \frac{\exp(-ipx_s)\psi(x_s)}{\tilde{\psi}(p)}, \quad (26)$$

in which $\tilde{\psi}(p)$ is the Fourier transform of $\psi(x)$. In the origin of the Fourier transform plane (i.e., $p = 0$), Eq. (26) simplifies to

$$\langle \hat{A} \rangle_w = \frac{\psi(x_s)}{\tilde{\psi}(0)}. \quad (27)$$

Since the denominator in Eq. (27) is a constant, it can be eliminated by normalizing the wave function. The weak value in this measurement system is proportional to the wave function at an arbitrary location $x_s$ of the silver plate. Thus, the light that passes through a pinhole set at the position of $p = 0$ is analyzed by taking the signal imbalance between two orthogonal polarization lights measured with two detectors. The real and imaginary parts of $\langle \hat{A} \rangle_w$ (i.e., the wave function) at the position $x_s$ of the silver plate are obtained by taking expected values of Pauli matrices $\hat{\sigma}_x$ and $\hat{\sigma}_y$, respectively:

$$\langle \psi_p | \hat{\sigma}_x | \psi_p \rangle = |\langle p | \psi \rangle|^2 \left(-2\theta \operatorname{Re} \langle \hat{A} \rangle_w \right), \quad (28)$$

$$\langle \psi_p | \hat{\sigma}_y | \psi_p \rangle = |\langle p | \psi \rangle|^2 \left(2\theta \operatorname{Im} \langle \hat{A} \rangle_w \right), \quad (29)$$

where $\hat{\sigma}_x = |H\rangle\langle V| + |V\rangle\langle H|$. The whole wave function can be reconstructed by scanning the silver plate in the object plane. Hence, this method determines the phase of the wave function by self-interference of light in mutually orthogonal polarization sates.

### B.  Operational formulation of the weak measurement of a wave function

To clarify the relationship between the method in Sec. IV A and the phase-retrieval method in Sec. II, we reformulate the experiment system in Fig. 3 in operator-based expressions [35] using another measurement method of the weak value. First, we obtain the field state of light passing through the polarizer $|S\rangle = (|H\rangle + |V\rangle)/\sqrt{2}$ placed at a position of $p$ in the Fourier-transform plane. After that, we measure the post-selection probability of the field state (i.e., the intensity of the transmitted light).

Then, the field state of light transmitted through the polarizer $|S\rangle$ can be written from Eqs. (24) and (25) as

$$\langle S|\psi_p\rangle = \langle p|\langle S|\exp(-i\,\theta\hat{A}\otimes\hat{\sigma}_y)|V\rangle|\psi\rangle$$

$$\cong \frac{1}{\sqrt{2}}\,\langle p|(1-\theta\hat{A})|\psi\rangle$$

$$= \frac{1}{\sqrt{2}}\,\langle p|\psi\rangle\left(1-\theta\langle\hat{A}\rangle_w\right). \quad (30)$$

Re-exponentiating Eq. (30), we find the following equation,

$$\langle p|\exp(-\theta\hat{A})|\psi\rangle = \langle p|\psi\rangle\exp\left(-\theta\langle\hat{A}\rangle_w\right). \quad (31)$$

The term on left-hand side of Eq. (31) is the post-selection probability amplitude obtained after transforming the wave function of the field state $|\psi\rangle$ by the attenuation operator $\exp(-\theta\hat{A})$ only at the location $x_s$, and the right-hand side term is the post-selection probability amplitude without such transformation, but modulated by the exponential form of the complex weak value. From the intensity measurement of the light, the relationship between two moduli $|\langle p|\exp(-\theta\hat{A})|\psi\rangle|$ and $|\langle p|\psi\rangle|$ can be derived as

$$\frac{|\langle p|\exp(-\theta\hat{A})|\psi\rangle|}{|\langle p|\psi\rangle|} = \exp\left(-\theta\mathrm{Re}\langle\hat{A}\rangle_w\right). \quad (32)$$

Thus, the ratio of the two moduli, which are the square root of the post-selection probabilities with and without the attenuation transform, corresponds to the exponential function of the real part of the weak value. In the same way, the relationship for the imaginary part of $\langle\hat{A}\rangle_w$ can be obtained from the intensity measurement of light passing through a polarizer with a quarter-wave ($\lambda/4$) plate [i.e., $|R\rangle = (|H\rangle + i|V\rangle)/\sqrt{2}$], instead of $|S\rangle$, as

$$\frac{|\langle p|\exp(-i\theta\hat{A})|\psi\rangle|}{|\langle p|\psi\rangle|} = \exp\left(\theta\mathrm{Im}\langle\hat{A}\rangle_w\right). \quad (33)$$

The measurements in Eqs. (28) and (32) or (29) and (33) are essentially equivalent, except for the treatment of the constant term on the right-hand side of Eq. (32) or (33), because the operators $\hat{\sigma}_x$ and $\hat{\sigma}_y$ in Eqs. (28) and (29) can be expressed in the forms $\hat{\sigma}_x = 2|S\rangle\langle S| - |H\rangle\langle H| - |V\rangle\langle V|$ and $\hat{\sigma}_y = 2|R\rangle\langle R| - |H\rangle\langle H| - |V\rangle\langle V|$, respectively. In other words, the measurement system for Eqs. (28) and (29) uses the signal imbalance detector between two orthogonal polarization lights, so the constant terms are automatically cut off.

It can be seen that Eqs. (32) and (33) have the similar in form to Eq. (6) in Sec. II. To make this point more clear, we rewrite Eqs. (32) and (33) from a quantum mechanical representation to a classical wave optics representation. The post-selection probability amplitude $\langle p|\exp(-\theta\hat{A})|\psi\rangle$ in the numerator of Eq. (32) can be rewritten as

$$\langle p|\exp(-\theta\hat{A})|\psi\rangle = \int \langle p|x\rangle\langle x|\exp(-\theta|x_s\rangle\langle x_s|)|\psi\rangle dx$$

$$\cong \int \exp(-ipx)\psi(x)[1 - \theta\delta(x - x_s)\psi(x)/\psi(x_s)]dx. \tag{34}$$

This equation indicates that the post-selection probability amplitude $\langle p|\exp(-\theta\hat{A})|\psi\rangle$ is the Fourier transform of the wave function $\psi(x)$ modulated by the function with an attenuation form $1 - \theta \cong \exp(-\theta)$ only at the point $x = x_s$ and not modulated otherwise. In the same manner, the numerator $\langle p|\exp(-i\theta\hat{A})|\psi\rangle$ in Eq. (33) can be rewritten as the Fourier transform of $\psi(x)$ modulated by the function with a unitary form $1 - i\theta \cong \exp(-i\theta)$ only at the point $x = x_s$ instead of $\exp(-\theta)$. Thus, this weak measurement method uses the rotation of polarization and the phase difference between orthogonal polarizations to add amplitude and phase modulation to the wave function to obtain its real and imaginary parts. In other words, this method is to recover the phase of the wave function by modulation filtering. Therefore, it can be seen that this method is similar to the phase-retrieval method by exponential filtering described in Sec.II.

**C. Quantum mechanical representation of the analytic phase retrieval method**

To make the similarity clearer, we will represent the phase-retrieval method in terms of a quantum mechanical description. The full initial state of light in the object plane in Sec. II is given by

$$|f\rangle = \int dx\, f(x)|x\rangle. \tag{35}$$

From Eq. (3), the field state of light passing through the exponential filter in the object plane is expressed at a transverse momentum $p$ in the Fourier-transform plane as follows,

$$\tilde{F}(p) = \exp(cs)F(p - ic)$$

$$= \exp(cs) \int dx\, f(x)\exp(-cx)\langle p|x\rangle$$

$$= \langle p|\exp[-c(\hat{X} - s\hat{I})]|f\rangle = \langle p|\exp(-ic\hat{A})|f\rangle, \tag{36}$$

where the interaction Hamiltonian for the exponential filtering is represented by $\hat{H}_I = c\hat{A} = -ic(\hat{X} - s\hat{I})$, in which $\hat{X}$ is the transverse position operator and $\hat{I}$ is the identity operator.

We therefore consider the case of weak measurements of phase retrieval using an exponential filter. Substituting Eq. (36) into Eq. (6) and using a first-order expansion approximation that the constant $c$ is small, Eq. (6) can be rewritten as

$$\frac{|\langle p|\exp(-ic\hat{A})|f\rangle|}{M(p)} = \exp\left\{c\left[s + \frac{d}{dp}\phi(p)\right]\right\}, \tag{37}$$

where $|M(p - ic)|$ and $\text{Im}\,\phi(p - ic)$ on the left-hand and right-hand sides, respectively, of Eq. (6) are approximated as follows,

$$|M(p - ic)| \cong \left|M(p) - ic\frac{d}{dp}M(p)\right| = \sqrt{M^2(p) + O(c^2)} \cong M(p), \tag{38}$$

$$\text{Im}\,\phi(p - ic) \cong \text{Im}\left[\phi(p) - ic\frac{d}{dp}\phi(p)\right] = -c\frac{d}{dp}\phi(p). \tag{39}$$

On the other hand, the weak value of the operator $\hat{A}$ in Eq. (36) is given by

$$\langle \hat{A} \rangle_w = \frac{\langle p|[-i(\hat{X} - s\hat{I})]|f\rangle}{\langle p|f\rangle}$$

$$= \frac{\frac{d}{dp}M(p)}{M(p)} + i\left[s + \frac{d}{dp}\phi(p)\right], \quad (40)$$

where $M(p) = |\langle p|f\rangle|$. Thus, Eq. (37) is rewritten from Eq. (40) as

$$\frac{|\langle p|\exp(-ic\hat{A})|f\rangle|}{|\langle p|f\rangle|} = \exp\left[c\mathrm{Im}\langle \hat{A}\rangle_w\right]. \quad (41)$$

This equation has the same form as Eq. (33) in the scanning system in Fig. 3, and shows that the ratio of their moduli (the square-root of the two post-selection probabilities) yields the imaginary part (i.e., the phase gradient of the wave function in the transverse momentum space) of the weak value in Eq. (40). Accordingly, this method using an exponential filter can be regarded as an extension of the weak measurement of the wave function. Although this method is not a direct method for obtaining the wave function at each coordinate point, it gives one analytical solution for the phase of the overall wave function without weak measurement approximation.

When using this phase retrieval method in practice, measurements are often made with the filter's slope parameter $c$ set to a small value (i.e., close to the weak measurement), in order to increase the signal-to-noise ratio. In addition, it is troublesome to prepare an exponential filter with an appropriate slope. To improve these defects, we have proposed a method to shift a Gaussian amplitude filter so that it has the same effect as an exponential filter [36]. Furthermore, we have presented a phase retrieval method using a scanning system with a Gaussian amplitude beam as a probe light [37], and then the experiments using the scanning phase retrieval have been performed to determine the modulus and phase of a two-dimensional object [38]. In the field of quantum optics, we have also proposed the method of reconstructing the wave function of a quantum state of light, in which the Gaussian wave function of the vacuum state is used as a probe for the analytic phase retrieval [39].

## V. Conclusions

The various interesting phenomena and results that appear when using quantum weak measurements are intrinsically related to the phase of the quantum system being measured. In this paper, we explained the effects of phase in two optical systems of weak measurements by means of the theory of analytic phase retrieval in classical wave optics. In the displacement measurement of polarized light in Sec. III, we showed that the pre- and post-selection of polarized light produces an exponential filtering effect that extends a function into the complex plane, resulting in anomalous displacement of the probe light as an effect on the phase of a zero in the complex plane. In the measurement of the wave function in Sec. IV, we presented that the pre- and post-selection of polarization by this system gives amplitude and phase modulation to the wave function, thereby obtaining the real and imaginary parts of the wave function. Therefore, this weak measurement system can be regarded as a filtering reconstruction method of the wave function, which is similar to the analytic phase retrieval using an exponential filter. The difference between these two methods is that the method based on weak measurements can directly measure the wave function at any coordinate point, while the exponential-filtering method can analytically reconstruct the overall wave function without the weak measurement approximation.

If weak measurements are made in two-state systems, such as spin-1/2 particles or polarized light, the results of this paper may be applicable regardless of whether the system is quantum or classical. This suggests that there is a theoretical crossover between weak measurements in quantum physics and analytic phase retrieval in classical wave optics via the physical quantity of the phase of the wave function. It is a future issue to investigate what extent the theory of analytic phase retrieval of wave functions is valid for measurements on various quantum systems.

## APPENDIX: DERIVATION OF EQ. (8)

According to the previous paper [32], we rewrite Eq. (6) as follows,

$$D(p',c) = -\text{Im}\,\phi(p'-ic)$$

$$= -\frac{1}{2i}[\phi(p'-ic) - \phi(p'+ic)], \quad (A1)$$

where $D(p',c) = \ln[|\tilde{F}(p')|/|M(p'-ic)|] - cs$ and the property that the phase is a real function [i.e., $\phi^*(p') = \phi(p')$] was used. The inverse Fourier transform of Eq. (A1) for the $p'$ coordinate can be written as

$$\mathfrak{I}^{-1}[D(p',c)] = -\frac{1}{4\pi i}\left[\int_{-\infty}^{\infty}\phi(p'-ic)\exp(ip'x)\,dp' - \int_{-\infty}^{\infty}\phi(p'+ic)\exp(ip'x)\,dp'\right]$$

$$= -\frac{1}{4\pi i}\int_{-\infty}^{\infty}\phi(p')[\exp(-cx) - \exp(cx)]\exp(ip'x)\,dp'$$

$$= \frac{-i\sinh(cx)}{2\pi}\int_{-\infty}^{\infty}\phi(p')\exp(ip'x)\,dp'$$

$$= -i\sinh(cx)\,\mathfrak{I}^{-1}[\phi(p')]. \quad (A2)$$

From the Fourier transform of Eq. (A2), we obtain a solution to the phase problem except for the origin of $x = 0$:

$$\phi(p) = \mathfrak{I}\left\{\frac{\mathfrak{I}^{-1}[D(p',c)]}{-i\sinh(cx)}\right\} \quad (x \neq 0), \quad (A3)$$

where $\mathfrak{I}\{\cdots\}$ denotes the Fourier transform for the $x$ coordinate.

The value of the function $\mathfrak{I}^{-1}[D(p',c)]$ at the origin of $x = 0$ is concerned with the inclination of a linear phase component of $\phi(p')$ for the $p'$ coordinate. The linear phase component can be represented by $\alpha p'$, where $\alpha$ is a tilt coefficient to be retrieved. Since the imaginary part of the linear phase $\alpha(p'-ic)$ expanded to the complex plane becomes a constant value $-\alpha c$, the tilt coefficient $\alpha$ can be obtained by dividing the value of $\mathfrak{I}^{-1}[D(p',c)]$ at the origin of $x = 0$ by the known constant $-c$. The linear phase component in the Fourier plane produces a shift in the object plane for the reconstructed object function. In practice, this effect can be ignored because the exact position of the object function is not so important in many applications. Thus, we here use Eq. (A3).

**Figure Captions**

Fig. 1. Schematic arrangement for coherent optical imaging. An object is illuminated by a coherent monochromatic plane wave. The wave front of the transmitted light through the object is Fourier transformed by lens $L_1$, and then the object image is obtained by lens $L_1$. $f_0$ is a focal length of lenses $L_1$ and $L_2$.

Fig. 2. Schematic diagram for weak measurements of the spin Hall effect of polarized light in a birefringent crystal. The probe beam $|\psi_0\rangle$ is pre-selected by the polarizer $P_1$ in the state $|S\rangle$, and the resulting beam $|\psi_i\rangle$ undergoes a unitary transformation by the birefringent crystal. After the output beam $|\psi_1\rangle$ from the crystal is post-selected by the polarizer $P_2$ in $|D_\theta\rangle$, the transverse displacement of the resulting beam $|\psi_f\rangle$ is measured by the CCD detector.

Fig. 3. Schematic diagram for the direct measurement of the transverse wave function by scanning a small half-wave ($\lambda/2$) plate. A one-dimensional object is illuminated by a stream of coherent photons with vertical polarization $|V\rangle$. The small half-wave plate rotates the polarization of the light passing through the object by $\theta$ in a small portion centered at position $x$. The wavefront in the object plane is then Fourier transformed by lens $L_1$. After being post-selected by the slit at a $p = 0$ in the Fourier plane, the polarized components of the collimated light by lens $L_2$ are measured to obtain the wave function at the position of the small half-wave plate.

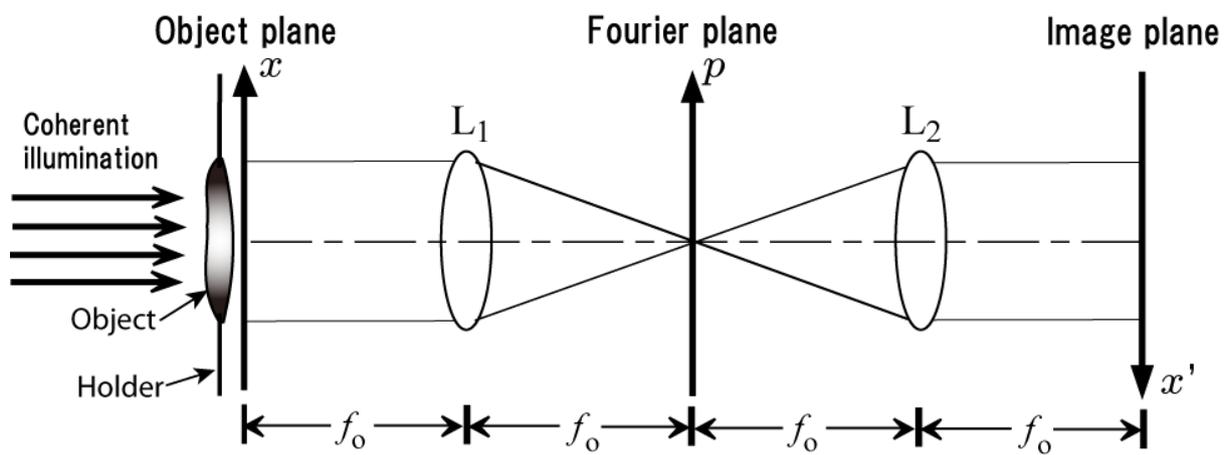

Fig. 1

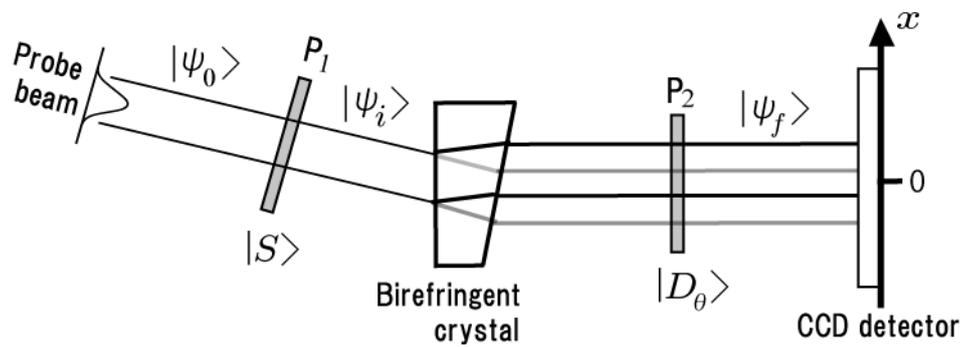

Fig. 2

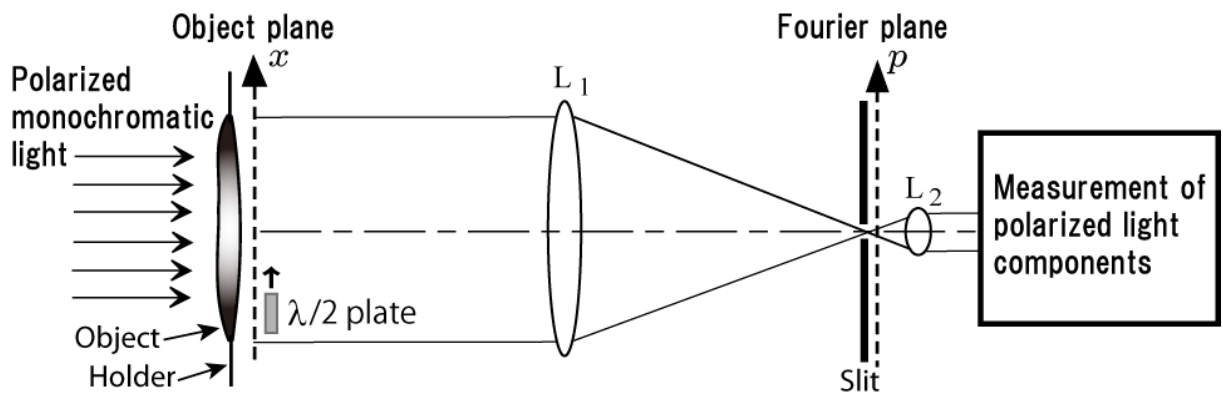

Fig. 3